\documentclass[sigconf]{acmart}

\usepackage[utf8]{inputenc}%
\usepackage{balance}
\usepackage{graphics}
\usepackage[T1]{fontenc}
\usepackage{txfonts}%
\usepackage{xcolor}%
\usepackage{booktabs}%
\usepackage{textcomp}%
\usepackage{balance}%
\usepackage{graphics}%
\usepackage{subcaption}%
\usepackage{microtype}
\usepackage[all=normal,floats,paragraphs]{savetrees}%
\usepackage{paralist}%
\usepackage{comment}%

\setcopyright{rightsretained}
\copyrightyear{2020}
\acmYear{2020}
\acmDOI{}
\acmConference[CHI '20 Workshop on Worker-Centered Design]{Workshop on Worker-Centered Design at ACM CHI '20}{April 25, 2020}{Honolulu, HI, USA}
\acmBooktitle{Workshop on Worker-Centered Design at ACM CHI '20,
  April 25, 2020, Honolulu, HI, USA}
\acmPrice{}
\acmISBN{}

\begin{document}

\title{What do crowd workers think about creative work?}

\author{Jonas Oppenlaender}
\email{jonas.oppenlaender@oulu.fi}
\affiliation{%
  \institution{University of Oulu}
  \city{Oulu}
  \country{Finland}
}

\author{Aku Visuri}
\email{aku.visuri@oulu.fi}
\affiliation{%
  \institution{University of Oulu}
  \city{Oulu}
  \country{Finland}
}

\author{Kristy Milland}
\email{kristy.milland@utoronto.ca}
\affiliation{%
  \institution{University of Toronto}
  \city{Toronto}
  \country{Canada}
}

\author{Panos Ipeirotis}
\email{panos@stern.nyu.edu}
\affiliation{%
  \institution{New York University}
  \city{New York, NY}
  \country{USA}
}

\author{Simo Hosio}
\email{simo.hosio@oulu.fi}
\affiliation{%
  \institution{University of Oulu}
  \city{Oulu}
  \country{Finland}
}

\renewcommand{\shortauthors}{Jonas Oppenlaender}

\begin{abstract}
Crowdsourcing platforms are a powerful and convenient means for
recruiting participants in online studies and
collecting data from the crowd.
As information work is being more and more automated by Machine Learning algorithms,
creativity~-- that is, a human's ability for divergent and convergent thinking~-- will play an increasingly important role on online crowdsourcing platforms.

However, we lack insights into 
what crowd workers think about creative work.
In studies in Human-Computer Interaction (HCI), the ability and willingness of the crowd to participate in creative work seems to be largely unquestioned.
Insights into the workers' perspective are rare, but important, as they may inform the design of studies with higher validity.
Given that creativity will play an increasingly important role in crowdsourcing,
it is imperative to develop an understanding of how workers perceive creative work.

In this paper, we summarize our recent worker-centered study of creative work on two general-purpose crowdsourcing platforms (Amazon Mechanical Turk and Prolific).
Our study illuminates 
what creative work is like for crowd workers on these two crowdsourcing platforms.
The work identifies several archetypal types of workers with different attitudes towards creative work, and discusses common pitfalls with creative work on crowdsourcing platforms.

\end{abstract}



\keywords{Crowdsourcing, creativity, crowd-powered creativity support tools}


\maketitle

\section{Introduction}%

Supporting creativity is considered a grand challenge in HCI~\cite{Shneiderman2009}.
Paid crowdsourcing platforms, such as Amazon Mechanical Turk and Prolific,
have become powerful and convenient means for 1) human subject recruitment in academic studies 
and 2) eliciting ideas.
These two attributes make the platforms attractive for \textit{crowd-powered creativity support tools}~\cite{DC2S2:2019,CCWorkshop:2019}~-- that is, online systems that aim to support the creativity of individuals or groups with help of the crowd. 

Workers on crowdsourcing platforms have become one of the most thoroughly studied sets of human subjects, 
and scientists have at their disposal a strong understanding of the demographics of workers on the crowdsourcing platforms (e.g., \citeN{p135-difallah.pdf,p16-ipeirotis.pdf,Paolacci2014}). 
The academic literature primarily investigates crowdsourcing from the perspective of the requester of work, with a focus on optimizing costs, response times, and work flows.
However, in studies in HCI, the ability and willingness of the crowd to contribute their creativity and participate in creative studies seems to be largely unquestioned.
Our understanding of how workers perceive creative work is limited. Insights into the workers' perspective are rare, but important, as they may inform the design of studies with higher validity.
Given that creativity will play an increasingly important role in crowdsourcing due to human's ability to excel in areas that machines fall short, such as divergent thinking, recombination, and analogical transfer,
it is imperative to develop an understanding of how crowd workers perceive creative work.


In a study conducted in August 2019 and published at CHI~'20~\cite{CHI20},
we investigated creative work from the workers' perspective on two commonly used crowdsourcing platforms that both compensate crowd workers for completing short tasks and participating in online surveys. 
We launched a worker-focused questionnaire in a task on these platforms, focusing on the workers' attitudes and preferences concerning creative work.

Our analysis of responses from 215~workers revealed
\begin{compactitem}
    \item differences between the workers of the two crowdsourcing platforms in both preferences and prior exposure to creative work,
    \item evidence for the nonna\"{i}vet\'{e} of crowd workers in regard to commonly used creativity tests, as a subset of creative work requested on the two crowdsourcing platforms,
    \item five different archetypes of crowd workers, based on different perceptions and attitudes towards creative work, and
    \item common pitfalls and recommendations for requesters of creative work on crowdsourcing platforms.
\end{compactitem}%
To the best of our knowledge, our work contributes the first worker-oriented study of creative work on two commonly used paid crowdsourcing platforms.
In the following, we describe the study and summarize our findings.

\section{The Study}

We launched a questionnaire as a task on two crowdsourcing platforms: Amazon Mechanical Turk and Prolific.
The two platforms are different in regard to worker demographics and what type of work is requested.
We selected the two platforms to provide us a complementary insight into creative work on two of the most commonly used crowdsourcing platforms.

The questionnaire included quantitative Likert-scale items as well as open-ended items asking the workers to elaborate on their answers.
The questionnaire was completed by 323 workers of which 215 (102~Amazon Mechanical Turk workers and 113~Prolific workers) met our qualification criteria (approval rate of greater than 98\%, 1000 completed tasks on Amazon Mechanical Turk, and prior encounters with creative work on the respective platform).
Workers were paid USD~1 on Amazon Mechanical Turk and GBP~1 on Prolific. Later, we noticed our estimated task completion times being too low and compensated workers with bonuses to pay at least minimum wage (USD~7.50) to all workers.

Three researchers open-coded all open-ended responses~\cite{braun2006.pdf}, allowing us to form a view of creative work on the platform and a sense for different archetypal worker profiles.

\section{Findings}

This section summarizes our main findings. The full paper, together with the questionnaire in the paper's Auxiliary Material, can be accessed on the ACM Digital Library~\cite{CHI20}.

\subsection*{Worker Archetypes}

Different types of workers may require different strategies for designing tasks when requesting creative work.

\textit{Professional workers} largely prefer to avoid creative work, as~-- in their view~-- this type of work is precarious and connected with considerable uncertainty in the amount of time required to complete the task and the reward may not be worth the effort in the view of this worker. Workers reported of unfair rejections for subjective tasks. Collaborative work flows, as in many experiments focusing on group creativity, may cause delays in completing the work. The professional worker will try to avoid such tasks, as they cause idle time and {forfeiting potential earnings from other tasks.}
Prior work found that professional workers complete the majority share of work on crowdsourcing platforms~(e.g., \cite{jdm14725.pdf}). Professional workers may thus have been exposed to a variety of different creative tasks on the crowdsourcing platforms, including standard psychometric tests, such as the Alternate Uses test. 
Foreknowledge and pre-exposure to such tasks may affect the validity of experiments and studies, as shown in the context of behavioral research~\cite{chandler2013.pdf}.

\textit{Pragmatic workers} may not care much about what type of tasks are given to them. In the view of this worker, requesters are going to request what they want, and some workers will be found who agree to work on the tasks. For this worker, it is important to have enough information about the task so that the worker can make an informed decision whether to accept the task or not.

\textit{Novelty seekers} are inspired and attracted by novel tasks.
The worker enjoys the unexpected and thus prefers working on creative tasks that the worker has never seen before.
This may lead to a ``tragedy of the commons'' effect, where the worker quickly gets to know the different tasks and then becomes bored with the platform.

\textit{Self-Developers} are less motivated by monetary rewards, but by opportunities for learning something new or about themselves.
Creative tasks, in this regard, are attractive to this worker, as they allow the worker to learn something about the worker's own personality.
For this type of worker, it is important to get post-task feedback in form of a debriefing with task results.

\textit{Casual workers} are perhaps most open to creative tasks, as there is a chance that this type of worker has never completed such tasks before. On the other hand, as casual workers work only few hours per week (or even per month) on the platform, the workers may never come in contact with creative work and may think that there is no creative work on the crowdsourcing platform at all.
This type of worker is also open to collaborative online experiments and workflows, as potentially precarious monetary rewards and task completion times do not play a strong role in the workers decision to start a task.%
%
%

\subsection*{Working collaboratively or Alone}

Supporting creative work has been regarded as one of HCI's grand challenges~\cite{Shneiderman2009}.
One trend in current research of the field is supporting the creativity of groups~\cite{p1235-frich.pdf}.

In our sample, we found clear evidence of workers
shunning on collaborative work.
Almost 90\% of the participants preferred solo work and only 23 participants (10.7\%) preferred collaborative work.
We isolated several reasons for this preference.
Enforced cooperation may cause issues both in time and rewards.
Cooperation issues were often brought up, with productivity and efficiency being the second most popular reason for workers to prefer solo work. 

Yet, there is a small, but clear, group of workers who enjoy collaborative creative work. For instance, one Prolific worker noted: ``The tasks that include others tend to be more exciting due to the anticipation of seeing how they will respond to each task.'' Other workers informally noted that they simply enjoy working together with other people -- a trait that, according to our study's findings, is 
shared by only a minority of workers on the two crowdsourcing platforms.

\subsection*{Pre-Exposure to Creative Work}

The existence of workers that are likely to have participated in academic studies has been documented in prior literature~\cite{chandler2013.pdf}.
Our study found evidence of worker nonna\"{i}vet\'{e} in regard to common creativity tests, as a subset of creative work requested on crowdsourcing platforms.
The past encounters raise a point to consider: Is the data collected from these participants on creativity tests {valid}?
Creativity tests measure a subject's creativity, but may also be used to measure other constructs. Lu et al., for instance, used the Alternate Uses test to measure unethical behavior \cite{8a01bb36375a507fa01ad95e2eb83ebf2f65.pdf}.
Creativity tests -- and especially those relying on an ``a-ha'' experience, such as Practical Insight Problems~-- 
may be affected by prior exposure.
If a relatively large portion of a study's sample has prior exposure to the very same creativity test, this should be considered in the study design and participant screening.

\subsection*{Learning from Creative Work}

A subset of the workers mentioned learning as a motivator for completing creative work. The proportion of participants reporting on some aspect of learning taking place during creative work on the crowdsourcing platform was high.
Of the 143 workers who elaborated on this question, only six (4.2\%) reported not learning anything at all. About 30\% of the participants (N = 43) mentioned that participation in creativity studies has led to discovering insights about their own personality. One fifth of the workers reported  either  developing  their  creative skills or awakening to the fact that they already are creative. Others mentioned having learned subject-specific skills from completing creative tasks. Increasing one’s  productivity and  concentrating for long periods of time were mentioned. Working on the platforms seems to encourage workers to engage and test their creativity in other areas, as exemplified by a worker who learned how to write scripts to support his work on the crowdsourcing platform.

%
%
%
\section{Discussion and Future Work}

Creative work on crowdsourcing platforms creates both challenges and
opportunities.

\subsection*{Challenge: Idea Ownership of Crowdsourced Ideas}
With creative work, issues with intellectual property and data use are likely to emerge~\cite{WolfsonLookBeforeLeap}~– even if currently ignored by the majority of requesters and platform operators.
If enterprises end up using creative ideas contributed by the crowd, what does this mean for patentable inventions?
The Participation Agreement of Amazon Mechanical Turk~\cite{PAMTurk} clearly states:
\begin{quote}
    \textit{[...] you (i) agree that all ownership rights, including all intellectual property rights, will vest with that Requester immediately upon your performance of those Tasks, and (ii) waive all moral or other proprietary rights that you may have in that work product. [...]}
\end{quote}
The potential in crowd-powered creativity is, it seems, well-matched by 
ethical concerns.
The reasonable way forward is to get workers who are interested in creative work 
to the same table with requesters, platform operators and policymakers to strike trade-offs and find common ground,
and to work towards incentive models that reward workers who donate their potentially valuable ideas for pennies on the dollar.

\subsection*{Challenge: New Incentive Models}
In organizations, contributions to organization-internal idea management systems are typically linked to an incentive if the contribution becomes a major hit.
In some countries, rewarding such contributions is required by law.
In Germany, for instance, ideas contributed to the company beyond the employee's contractual agreements are to be awarded with a share of the profit that the idea brings to the company. 
Crowdsourcing platforms, on the other hand, avoid contractual agreements with their workers. Workers are freelancers and are not entitled to any kind of benefits.
To address the challenge of intellectual property rights on crowdsourcing platforms, new types of incentive models are needed for creative work. 
At the same time, the situation could call for some regulatory actions targeted to improve the workers' rights. Yet, this is extremely challenging from a legal perspective, since the platform is global and workers from all over the world, with differing backgrounds and from different cultural expectations of work in general.

\subsection*{Opportunity: Choosing the Right Platform}
What crowdsourcing platform should be used for eliciting creative work and studying creativity?
Our work found differences in how workers perceive creative work. Professional workers may have a negative attitude towards creative work, as it may take more time to complete and may be associated~-- in the view of the worker~-- with considerable uncertainty in regard to the rewards.
Our data indicates that casual workers on Prolific have been less exposed to creativity studies, show more interest in creative work, and may therefore be a better participant pool for creativity-oriented research.

\subsection*{Opportunity: Dedicated Platforms for Creativity}
Prior literature suggested tailored platforms for human-centered experiments~\cite{Gadiraju2-1wgg9voexzjhw3.pdf}.
Future tailored platforms could particularly target and cater to participants who enjoy {creative microtask work}.
Alternatively, recruitment tools interfacing with current crowdsourcing platforms~-- similar to TurkPrime~\cite{TurkPrime}~-- could be created to help with cultivating a participant pool interested in creativity studies. 
To this end, a more complete understanding of the design space for creative work on crowdsourcing platforms is needed.


\subsection*{How do crowd workers define creativity?}%
Prior research on creativity found over three quarters of the recent publications in HCI do not offer a definition or clarification of the term creativity~\cite{p1235-frich.pdf}.
Given that empirically based contributions are prevalent in creativity studies in HCI~\cite{p1235-frich.pdf}, it is imperative to develop an understanding of how all stakeholders of creative work conceive creativity.
A {conceptual gap} between how workers conceive creativity and how requesters (often implicitly) define creativity in their studies may potentially impact the validity of a creativity study. Ideally, this conceptual gap should be minimized.

Future studies could investigate how crowd workers define and understand creativity. This would provide the workers' perspective for requesters to consider when using crowdsourcing platforms for creative work and creativity-related experiments.

\section{Conclusion}

\balance{}

Crowdsourcing platforms are excellent sources for recruiting participants for creativity-oriented research and for eliciting original thought.
Most likely we have not yet unlocked the full potential of this combination, and there is much to discover about the feasibility of crowdsourced creativity for both scientific and industry purposes.

With our study, we wish to raise awareness of some of the shortcomings of the current research and practice in using crowdsourcing platforms for creative work. 
%
%
%
Distinct calls in HCI have been made for researchers to 
contribute toward developing an improved and inclusive society~\cite{Bardzell:2010:FHT:1753326.1753521}. This includes the workers on platforms who we now commonly engage as participants in online studies or for harvesting data. Indeed, we can regard crowd work as inherently participatory: The labor obtained from crowdsourcing platforms is not purchased as cycles of human-computational labor produced by anonymous ``humans-as-a-service''~\cite{Irani:2013}, but it originates from stakeholders, even if paid ones, in a co-creation process. Acknowledging the needs and wishes of the workers allows us to transition to a more humane relationship with the workers.






\balance{}
\bibliographystyle{ACM-Reference-Format}
\bibliography{paper}

\end{document}